\documentclass[prl,twocolumn,floatfix,showpacs,superscriptaddress]{revtex4}
\usepackage{graphicx}
\usepackage{amsmath}
\begin{document}
\newcommand{\vf}{v_{\rm F}}

\title{Transport regimes in surface disordered graphene sheets}
\author{E. Louis}
\affiliation{Departamento de F\'{\i}sica Aplicada, Unidad Asociada del
Consejo Superior de Investigaciones Cient\'{\i}ficas and
Instituto Universitario de Materiales, Universidad
de Alicante, San Vicente del Raspeig, Alicante 03690, Spain.}
\author{J. A. Verg\'es}
\affiliation{Departamento de Teor\'{\i}a de la Materia Condensada,
Instituto de Ciencias de Materiales de Madrid (CSIC),
Cantoblanco, Madrid 28049, Spain}
\author{F. Guinea}
\affiliation{Departamento de Teor\'{\i}a de la Materia Condensada,
Instituto de Ciencias de Materiales de Madrid (CSIC),
Cantoblanco, Madrid 28049, Spain}
\author{G. Chiappe}
\affiliation{Departamento de F\'{\i}sica Aplicada, Unidad Asociada del
Consejo Superior de Investigaciones Cient\'{\i}ficas and
Instituto Universitario de Materiales, Universidad
de Alicante, San Vicente del Raspeig, Alicante 03690, Spain.}
\affiliation{Departamento de F\'{\i}sica J.J. Giambiagi, Facultad de
Ciencias
Exactas, Universidad de Buenos Aires, Ciudad Universitaria,
1428 Buenos Aires, Argentina.}

\date{\today}

\begin{abstract}
We investigate the size scaling of the conductance of surface
disordered graphene sheets of width $W$ and length $L$. Metallic
leads are attached to the sample ends across its width.
At the Dirac point, $E = 0$, the conductance
scales with the system size as follows:
i) For constant $W/L$, it remains constant as size is increased,
at a value which depends almost linearly on that ratio
%, i.e., a behavior close to diffusive transport
; this scaling  allows the definition
of a conductivity value that results similar to the experimental one.
ii) For fixed width, the conductance decreases exponentially with length $L$,
both for ordered and disordered samples.
Disorder reduces the exponential decay, leading to a higher conductance.
iii) For constant length, conductance increases linearly with width $W$,
a result that is exclusively due to the tails of the states of the
metallic wide contact.
iv) The average conductance does not show an appreciable dependence on
magnetic field until fields such that the fluc per unit cell approaches the
quantum unit.
Away from $E = 0$, the conductance shows the behavior expected in 
two-dimensional systems with surface disorder, i.e., ballistic transport.
%These results highlight the role played by the
%evanescent waves originating at the metallic leads in determining the apparent
%diffusive behavior of graphene at $E \approx 0$ while showing a more
%standard detailed scaling in width and length.
\end{abstract}


\maketitle

\noindent {\it Introduction.} The electronic transport in atomically thin
graphene samples is a subject of great current 
interest \cite{Netal04,Betal04,Betal05,Netal05,Netal05b,Zetal05b,Metal05c}.
The scaling with the sample dimensions  \cite{Netal05} suggest a
diffusive behavior, with a universal conductivity at the lowest carrier
concentrations \cite{Netal05,Zetal05b}. 
The limit of low concentrations is difficult to analyze theoretically, as the
Fermi wavelength becomes comparable to the separation between scatterers, and
even to the sample size. An analysis based on the Born
approximation \cite{PGN06} leads to a universal conductivity at low
temperatures, although its value is somewhat smaller than the one observed
experimentally. The approximations involved in this approach, however, are
expected to fail at the lowest concentrations.
Field theoretical arguments \cite{AE06,A06,OGM06} suggest the existence
of a localized regime in the limit of zero temperature
and zero carrier concentration.
At zero doping, clean graphene systems show an unusual
scaling of the conductance on sample size, consistent with diffusive
behavior \cite{TTTRB06}. This pseudo-diffusive behavior has also been found
in SNS junctions \cite{TB06} and graphene bilayers \cite{SB06}.

In this work, we numerically study the electronic transport in surface
disordered graphene sheets both at finite dopings
and in the limit of zero carrier concentration.
As bulk disorder in graphene sheets seems to be rather low,
we focus on the effects of rough edges,
with disorder concentrated at the surface of the system. 
Our results show that the pseudo-diffusive regime identified
in \cite{TTTRB06} persists in the presence of disorder, namely, near
the band center the conductance is proportional to the sheet width and
almost inversely proportional to its length.
Although our results fit apparently the requirements of diffusive scaling
in 2D, a closer look reveals important differences.
Certainly, when plotting the conductance of samples of fixed width
as a function of the sample length,
an exponential decrease is obtained, that is, the standard result
for a quasi 1D system with any kind of disorder.
The remarkable thing in graphene is that this exponential decrease
survives in ordered samples. Actually, the presence of disorder slows
down the exponential decay. These results can be rationalized in terms
of transmission mediated by evanescent waves
generated at the metallic leads. On the other hand,
the conductance for fixed length is proportional to the
sample width at all energies, a behavior
that does not distinguish between diffusive and ballistic regimes.
Finally, for  sufficiently high carrier concentrations, the conductance shows 
the ballistic behavior expected in 2D systems with surface disorder, 
namely, increase with the system size for constant $W/L$, linear increase
with $W$ for constant $L$, and exponential decrease with $L$ for constant $W$.
\begin{figure}
\includegraphics[width=0.7\columnwidth]{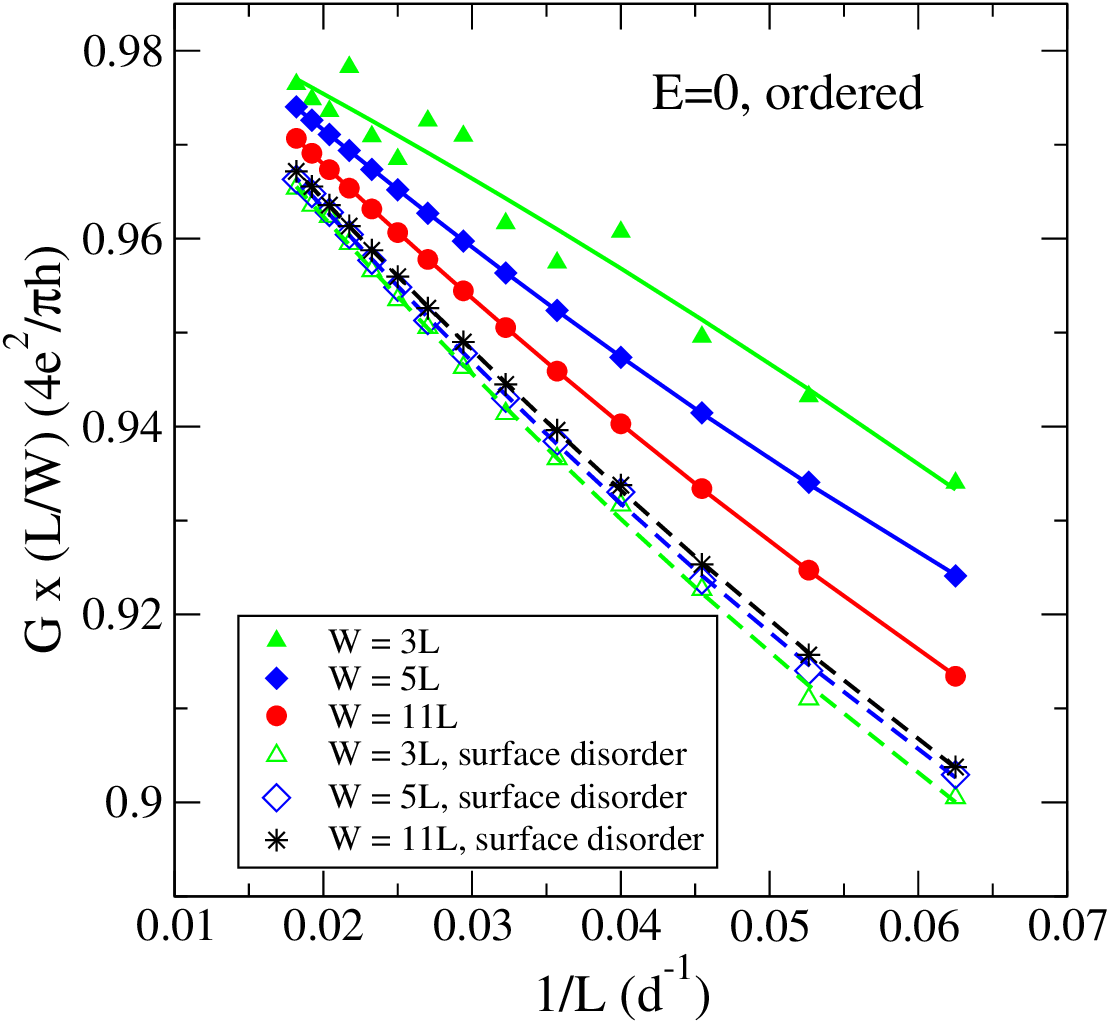}
\caption{(Color online) Scaling of the conductance of stripes of different
  widths, $W$, as function of length, $L$. 
}
\label{scaling_ordered}
\end{figure}

\noindent {\it Methods: graphene samples and conductance calculations.}
We describe the valence and conduction bands of graphene by a tight binding
hamiltonian with nearest neighbor hoppings only:

\begin{equation}
{\cal H} = t \sum_{i,j} c^\dag_i c_j + h. c.
\end{equation}
where sites $i$ and $j$ denote the nearest neighbor nodes in the honeycomb
lattice. 
The low energy electronic spectrum,
$| \epsilon_{{\bf \vec{k}}} | \ll t$, can be approximated by the Dirac
equation: 
\begin{equation}
\epsilon_{{\bf \vec{k}}} \approx \pm \vf | {\bf \vec{k}} |
\end{equation}

where $\vf = ( 3 t d ) / 2$, and $d$ is the distance between sites in the
honeycomb lattice.

Samples with surface disorder were produced by randomly removing sites at
the sheet edges. 
%A typical example is shown in the upper panel of
%Fig. \ref{fig:sample}.
The leads were simulated by a purely imaginary selfenergy
independent of energy, that was attached across the sample width.
%(thicker circles of upper panel in Fig. \ref{fig:sample}). 
%In order to help
In the interpretation of the numerical results for surface
disordered sheets, perfect samples without and with Anderson disorder
at its edges \cite{CL96,LC97} were also investigated . The
latter was introduced by randomly sorting the orbital energies
at the surface sites within the range $[-\Delta,
\Delta]$. 

The conductance was calculated by means of 
an efficient implementation of Kubo's formalism \cite{Ve99}.
For a current propagating in the $x$--direction,
the static electrical conductivity is given by:
\begin{equation}
{\mathcal G} = -2 {\left( \frac{e^2}{h} \right)} 
{{\rm Tr} \left [(\hbar {\hat v_x})
{\rm Im\,}{\widehat G}(E)(\hbar {\hat v_x})
{\rm Im\,} {\widehat G}(E)\right ]} \;,
\end{equation}
where the velocity (current) operator ${\hat v_x}$ is related
to the position operator ${\hat x}$ through the equation of motion
$\hbar {\hat v}_x = \left [ {\widehat H},{\hat x} \right ]$,
$\widehat{H}$ being the Hamiltonian.  ${\widehat G}(E)$ is the Green function
of the system with the leads already incorporated. All results include the
spin degeneracy.

\noindent {\it Results.} 
Fig.[\ref{scaling_ordered}] shows typical results for stripes
without and with disorder. At $E=0$, the scaling of the conductance in clean samples as
$G = 4 e^2 / ( \hbar \pi ) \times W / L$\cite{TTTRB06} is already obtained with high
accuracy in not too wide samples. At higher energies, $E=0.5$, the
conductance becomes ballistic.
This pseudodiffusive regime, was already analyzed in
clean systems in \cite{TTTRB06}. In a clean square system, the incoming
channels can be characterized by the transverse momentum, $k_y$. The
electronic spectrum of a graphene stripe at finite transverse momentum shows
a gap for $- \vf | k_y | \le \epsilon \le \vf | k_y |$. Hence, states with
transverse momentum $k_y$ decay away from the boundaries as $e^{- | k_y | x}$,
and lead to a transmission $T_{k_y} \propto e^{- 2 | k_y | L}$, where
$L$ is the length of the system.  The number of
channels scale as the width of the system, $W$.
In sufficiently large systems, the sum
over channels can be replaced by an integral over $k_y$ leading to a
conductance $G$ which scales as $G \propto W L^{-1}$. Away from
$E=0$ the conductance increases linearly with the system size.
This is the expected ballistic behavior
of a quantum billiard with either surface disorder or with an amount
of defects proportional to $L$ (a defect concentration decreasing as $1/L$) 
\cite{CL96,VL99}.
%Fig. \ref{fig:sample} shows the conductance versus
%energy in a graphene sample representative of those investigated in this work.
%The conductance remains almost constant over a narrow range around $E=0$.
%The actual value of ${\mathcal G}$ in that region depends on disorder
%and on the shape and size of the sample (see below). 

\begin{figure}
\includegraphics[height=10cm,width=0.9\columnwidth]{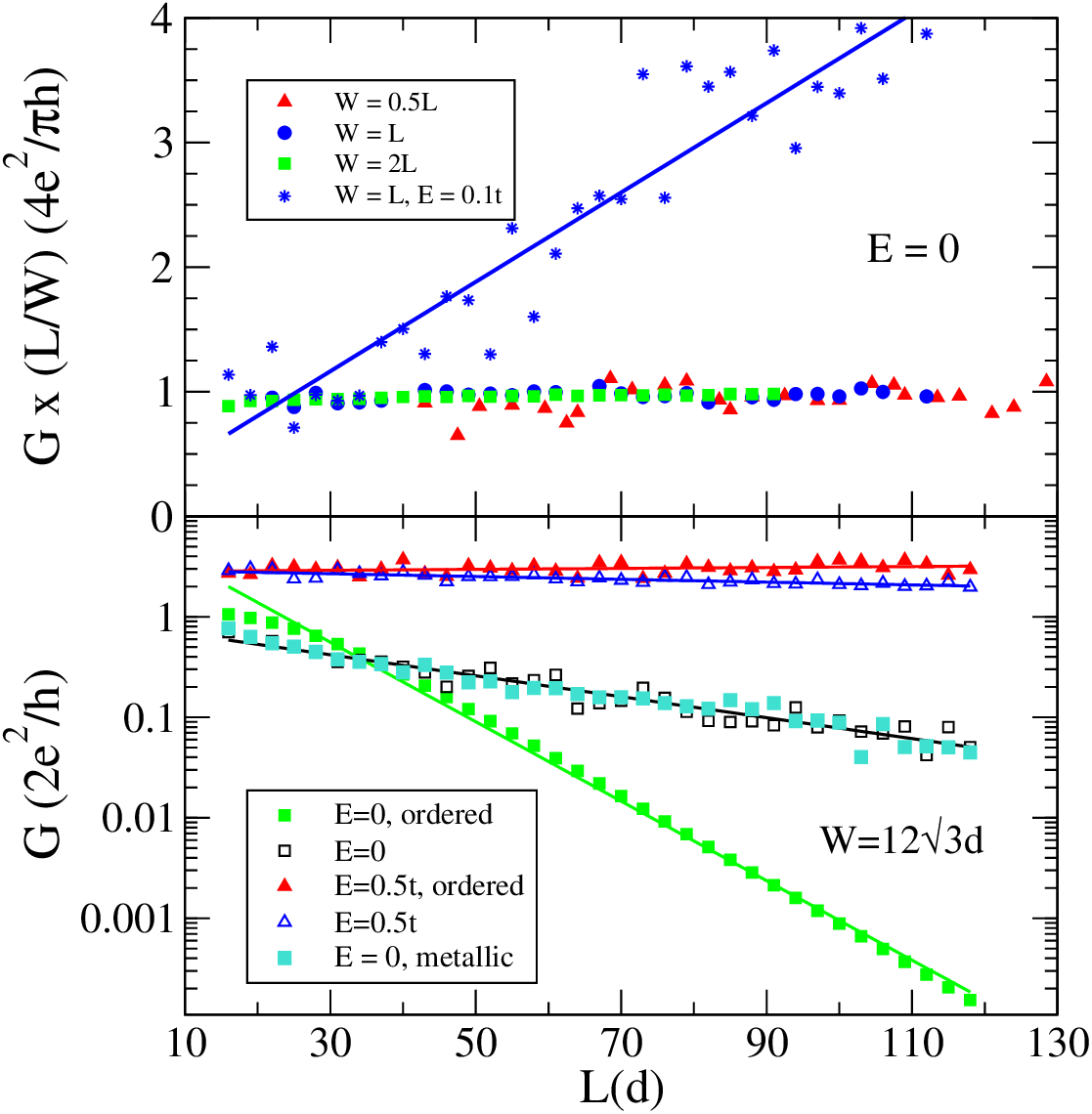}
\caption{(Color online)
Top panel: Scaling with length $L$ of the conductance in samples with
width over length ratios and surface disorder. Bottom panel. Scaling with
length of stripes of fixed width. The points labelled metallic correspond to
a stripe with a subband crossing at $E=0$. All other samples have finite size
gaps near $E=0$.} 
\label{fig:square}
\end{figure}

The scaling of the conductance
with stripe length in systems with surface disorder and different
$W/L$ ratios is shown in the upper panel of Fig.[\ref{fig:square}]. The
conductance 
at $E=0.1$ deviates very slightly from ballistic behavior. Disordered samples
show a length independent conductivity close to that estimated analytically
in the clean linmit\cite{TTTRB06}, and this regime is attained even for
widths smaller than the length. The conductance scales exponentially with
$L$. In a clean systems, this behavior arises from the existence of minigaps
separating subbands with well defined periodicity in the transverse
direction. The corresponding decay length shorter in clean samples than in
disordered samples, leading to an enhancement of the conductance in
semiconducting disordered samples at $E=0$. 

\begin{figure}
\includegraphics[width=0.8\columnwidth]{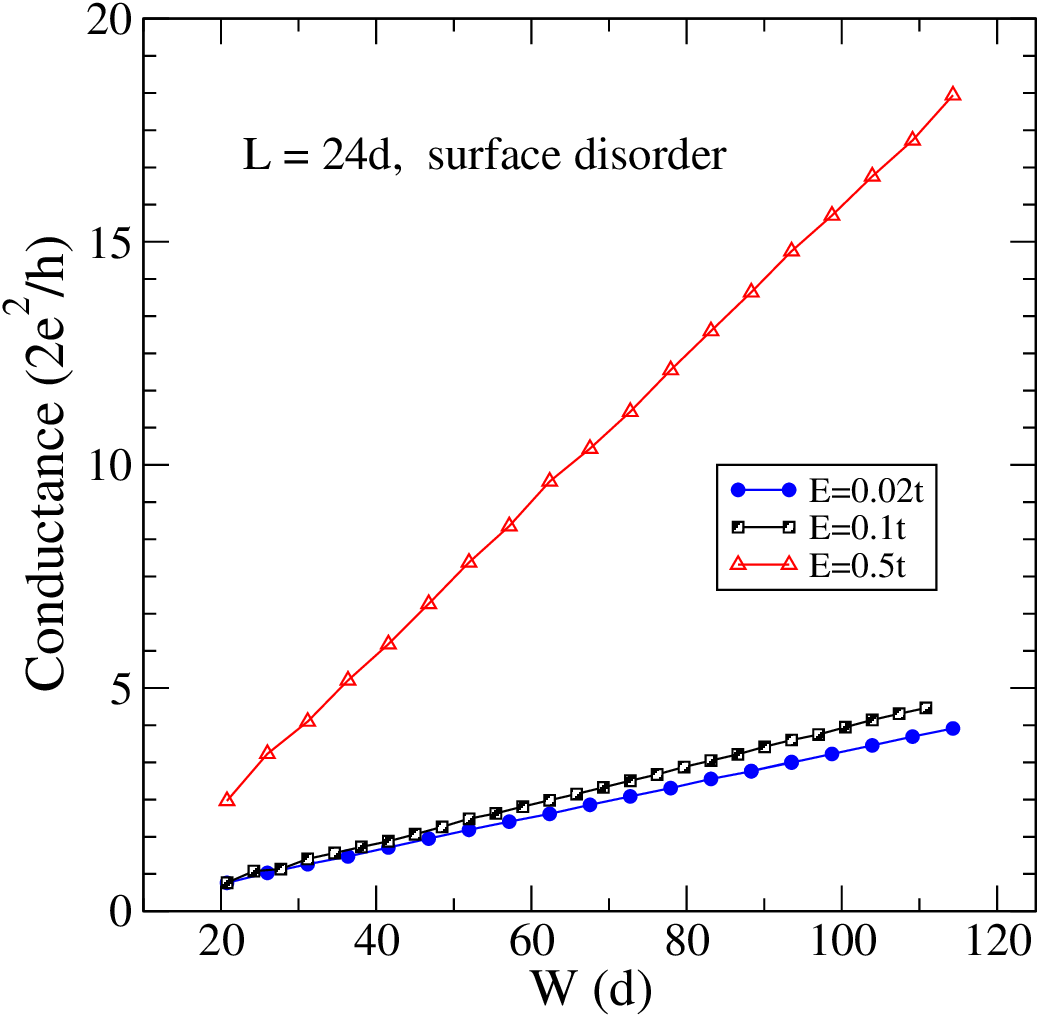}
\caption{(Color online)
Conductance (in units of the conductance quantum) through surface disordered
graphene samples of size $W \times 24d$ versus the sample width $W$ 
(in units of the C-C distance $a$). Results for three energies are shown.
\label{fig:stripe-W}}
\end{figure}

The enhancement of the conductance in the localized regime in disordered
samples is probably related to the formation of resonances at $E=0$ near
defects. The existence of these resonances has been well established at
edges\cite{WS00,W01}, cracks\cite{Vetal05}, and
vacancies\cite{Petal05}. After the first version of this paper was posted,
related resonances where discussed in the continuum limit\cite{T06}. Note
that, in addition to $E=0$ resonances induced by disorder, long wavelength
modulations of the chemical potential will move one of the edges of the gap at
fixed parallel momentum towards $E=0$, reducing the decay length and
enhancing the conductance.

%As shown in Fig.[\ref{fig:square}], disorder seems to slow down
%the decay of the conductance with the system length. 
%In order to reinforce this result we have also
%calculated the conductance through
%a geometrically perfect sample with Anderson disorder.
%The results shown in the same Figure
%are in line with those found for surface disordered samples:
%we conclude that surface disorder of any kind facilitates transport. 
%This counterintuitive result can be understood in terms of
%the following simple reasoning.
%For a given parallel momentum $k_y$, a smooth local shift in the
%potential, $V ( x )$, induces a local gap in the energy range $- \vf | k_y |
%+ V ( x ) \le \epsilon \le \vf | k_y | + V ( x )$. Thus, incoming electronic
%states with energy $\epsilon = 0$ and momentum $| k_y | \le V ( x ) / \vf$ are
%propagating waves in these regions. The states with $\vf | k_y | > V ( x )$
%are described by evanescent waves, but the decay length becomes $\kappa ( x )
%\approx \vf^{-1} {\rm min}
%\left( | V ( x ) + \vf k_y | , | V ( x ) - \vf k_y | 
%\right)$. Hence, the transmission is enhanced for all momenta,
%irrespective of the sign of $V ( x )$. 
%These arguments can be extended to the case of a
%graphene bilayer at $E=0$\cite{SB06} where the decay length 
%must also satisfy $\kappa = k_y$.

The scaling of the conductance with the width of a sheet of constant length is
shown in Fig. \ref{fig:stripe-W}.
The numerical results clearly indicate that the
conductance increases linearly with the sample width with
a slope that depends on the concentration of carriers (or the energy).
This  result is characteristic of both ballistic and
diffusive behaviors in 2D, and cannot therefore be used
to discriminate the transport regime in this case.
A remarkable feature of the results shown in Fig. \ref{fig:stripe-W}
is its very low dispersion.
This could be understood by noting that the increase
in conductance is exclusively due to an increasing number of metallic tails
through the bulk of the graphene sheet,
and therefore weakly sensitive to surface disorder.

\begin{figure}
\includegraphics[width=0.8\columnwidth]{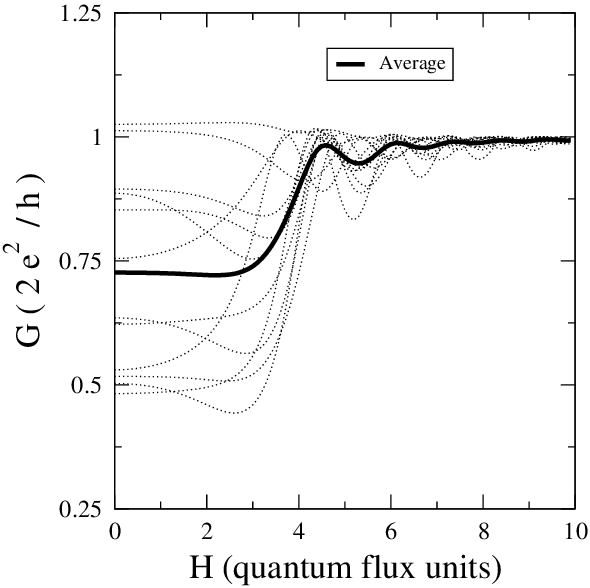}
\caption{(Color online)
Conductance (in units of the conductance quantum) as function of the total
magnetic flux through the graphene lattice in $40 d \times 40 d$ clusters
and $E= 0.02$, with disorder at the edges.
A magnetic field of 1 Tesla corresponds,
approximately, to 0.04 flux units through the cluster.}
\label{fig:flux}
\end{figure}

We show in Fig.[\ref{fig:flux}] the dependence of the conductance on magnetic
field 
for different disorder realizations \cite{LV01}. 
The area of the sample, $40 d \times 40 d$, and $62$ unit cells, is such that
one flux unit 
through it is equivalent to approximately 660 Teslas. Hence, the
magnetoresistance for fields attainable in the laboratory is negligible.
This result is consistent with semianalytical calculations using a continuum
model in the clean limit\cite{PSWG06}. 
The magneto-resistance shows oscillations when the flux per unit
cell is of order 0.05. At very high fields, the conductance becomes of order
of one quantum unit, and it shows no dependence on disorder realization.
Note that, in the regime studied here, the quasiclassical arguments used when discussing 
either weak localization or weak antilocalization effects in graphene cannot
be used\cite{Metal06,MG06,Metal06c}.

\begin{figure}
\includegraphics[width=0.6\columnwidth]{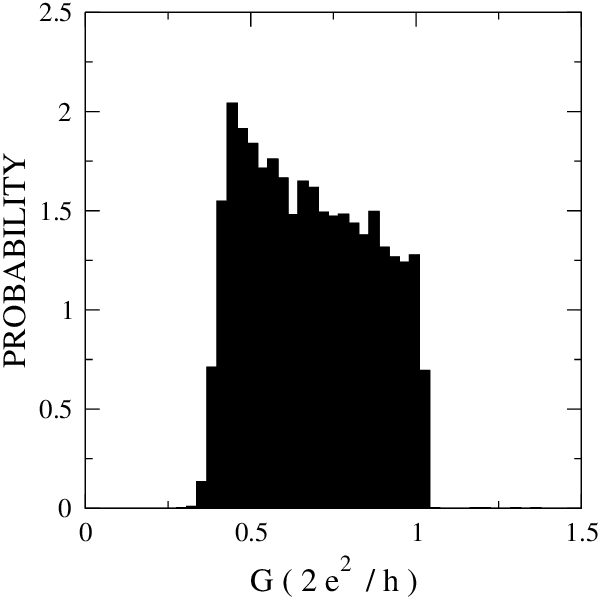}
\caption{
Conductance distribution obtained at $E=0.001t$ for ia set of 10,000 graphene
randomly generated samples of size $24d \times 24d$. Surface disorder
is restricted to a $4d$ fringe around the sample surface. Good
metallic contacts are attached at opposite sides of the sample.
\label{hist}}
\end{figure}

The study of the conductance distribution for samples of an
approximate square shape at an energy $E=0.001t$ very close to the Dirac
point further explains the role played by the metallic tails in the
conductance behavior. Fig. \ref{hist} shows that conductance is 
larger than $\approx 0.37 (2 e^2/h)$ for this geometry
(the tails contribution that is minimally
affected by surface disorder) and fluctuates below 1 as it does in a
standard quantum billiard in the case of point contacts. The existence of an
abrupt upper cutoff resembles the case studied in\cite{MWGG03}. Nevertheless,
a significant difference is clear; while the upper conductance limit is due
to the incidence of only one channel in the billiard case, it is due
to the intrinsic small number of channels (0 or 1) of graphene near the
band-center.

\noindent {\it Concluding Remarks}.
The numerical calculations of the conductance through surface
disordered graphene sheets presented in this work reproduce
the quasi-diffusive behavior found by other authors in 
ordered graphene at the Dirac point\cite{TTTRB06,B06,TB06,SB06}. Specifically,
the conductance remains constant when the size of the system is increased,
as opposed to the linear increase with the system size found
at any other energy. However, we found this behavior only when
the size of the system is increased in such a way that the width
to length ratio is kept constant. A value of $\approx 0.75 (2 e^2/h)$
is obtained for this pseudo-conductivity, not that far
from the experimental one.

The conductance of the stripes calculated here changes qualitatively as the
aspect ratio of the sample is varied. In stripes where the length is much
larger that the width, the pseudodiffusive behavior described above is replaced
by an exponential decay with length, a sign of localization. In this regime,
disorder leads to longer decay lengths, probably due to the formation of
resonances at the Dirac point, $E=0$.

 Finally, we have also shown that a magnetic field with a magnetic length
 much larger than the lattice spacing does not change
appreciably the conductance near the Dirac point.

{\em Acknowledgments}
Financial support by the Spanish MCYT (grants FIS200402356,
MAT2005-07369-C03 and NAN2004-09183-C10-08),  
the Universidad de Alicante, the Generalitat Valenciana
(grant GRUPOS03/092 and grant GV05/152), the Universidad de Buenos
Aires (grant UBACYT x115) and the Argentinian CONICET
is gratefully acknowledged. 
GC is thankful to the Spanish "Ministerio de Educaci\'on y Ciencia"
for a Ram\'on y Cajal grant. F. G. acknowledges
funding from MEC (Spain) through grant FIS2005-05478-C02-01
and the European Union Contract 12881 (NEST), and the Comunidad de Madrid, through
the program CITECNOMIK, CM2006-S-0505-ESP-0337.

\bibliography{prl}

\end{document}